# Deep generative models of genetic variation capture mutation effects

**Adam J. Riesselman**[*]
Program in Biomedical Informatics
Harvard Medical School
`ariesselman@g.harvard.edu`

**John B. Ingraham***
Program in Systems Biology
Harvard University
`ingraham@fas.harvard.edu`

**Debora S. Marks**
Department of Systems Biology
Harvard Medical School
`debbie@hms.harvard.edu`

* Equal contribution

## Abstract

The functions of proteins and RNAs are determined by a myriad of interactions between their constituent residues, but most quantitative models of how molecular phenotype depends on genotype must approximate this by simple additive effects. While recent models have relaxed this constraint to also account for pairwise interactions, these approaches do not provide a tractable path towards modeling higher-order dependencies. Here, we show how latent variable models with nonlinear dependencies can be applied to capture beyond-pairwise constraints in biomolecules. We present a new probabilistic model for sequence families, DeepSequence, that can predict the effects of mutations across a variety of deep mutational scanning experiments significantly better than site independent or pairwise models that are based on the same evolutionary data. The model, learned in an unsupervised manner solely from sequence information, is grounded with biologically motivated priors, reveals latent organization of sequence families, and can be used to extrapolate to new parts of sequence space.

## Introduction

Modern medicine and biotechnology are routinely challenged to both interpret and exploit how mutations will affect biomolecules. From interpreting which genetic variants in humans underlie disease, to developing modified proteins that have useful properties, to synthesizing large molecular libraries that are enriched with functional sequences, there is need to be able to rapidly assess whether a given mutation to a protein or RNA will disrupt its function [1, 2]. Motivated by these diverse applications, new technologies have emerged that simultaneously assess the effects of thousands of mutations in parallel [3-25] (sometimes referred to as "deep mutational scans"[26] or 'MAVEs'[27, 28] ). In these assays, the measured attributes range from ligand binding, splicing and catalysis [4, 8, 11, 13, 21, 23, 29] to cellular or organismal fitness under selection pressure [5-7, 9, 12, 14, 17, 19, 20].

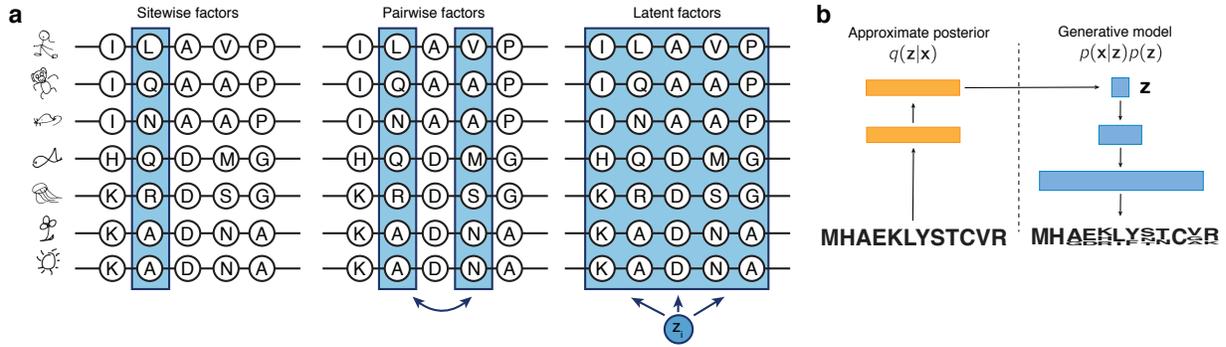

**Figure 1. A nonlinear latent variable model captures higher-order dependencies in proteins and RNAs. a.** *In contrast to sitewise and pairwise models that factorize dependency in sequence families with low-order terms, a nonlinear latent variable model posits hidden variables z that can jointly influence many positions at the same time.* **b.** *The dependency p(x|z) of the sequence x on the latent variables z is modeled by a neural network, and inference and learning is made tractable by jointly training with an approximate inference network q(z|x). This combination of model and inference is also known as a variational autoencoder*

Since sequence space is exponentially large and experiments are resource-intensive, accurate computational methods are an important component for high-throughput sequence annotation and design. Many computational tools have been developed for predicting the effects of mutations, and most progress in efficacy of predictions has been driven by the ability of models to leverage the signal of evolutionary conservation among related sequences [30-35]. While previous approaches analyzed this signal in a residue-independent manner, recent work has demonstrated that incorporating inter-site dependencies using a pairwise model can power state of art predictions for high-throughput mutational experiments [36-38]. Although this incorporation of pairwise epistasis represented an important step forward, contemporary models based on natural sequence variation are still unable to model higher-order effects. This is despite the frequent observation that higher order epistasis pervades the evolution of proteins and RNAs [39-42]. Naïvely, one way to address this would be to simply extend the pairwise models with third or higher terms, but this is statistically unfeasible: fully-parameterized extensions of the pairwise models to third-order interactions will already have approximately $\sim 10^9$ interaction terms for a protein of length only 200 amino acids. Even if such a model could be engineered or coarse-grained [43] to be computationally and statistically tractable, it will only marginally improve the fraction of higher-order terms considered, leaving $4^{th}$ and higher order interactions seemingly out of reach.

The intractability of higher order interaction models for modeling epistasis in proteins is a consequence of how these models describe data: in general, every possible higher order interaction requires explicit incorporation of a unique free parameter that must be estimated. However, this is not the only way to model higher-order correlations. Rather than describing them by explicit inclusion of parameters for each type interaction, it is possible to instead implicitly capture higher-order correlations by means of latent variables. Latent variables models posit hidden factors of variation that explain observed data and involve joint estimation of both hidden variables for each data point as well as global parameters describing how these hidden variables affect the observed. Two widely used models for the analysis of genetic data, PCA and admixture analysis [44-46] can be cast as latent variable models with linear dependencies. Although these linear latent variable models are restricted in the types of correlations that they can model, replacing their linear dependencies with flexible nonlinear transformations can in principle allow the models to capture arbitrary order correlations between observed variables. Recent advances in approximate inference [47, 48] have made such nonlinear latent variable models tractable for modeling complex distributions for many kinds of data, including text, audio, and even chemical structures [49], but their application to genetic data remains in its infancy.

Here, we develop nonlinear latent variable models for biological sequence families and leverage approximate inference techniques to infer them from large multiple sequence alignments. We show how a Bayesian deep latent variable model for protein sequence families can be used to predict the effects of mutations and organize sequence information, all while being grounded with biologically motivated architecture and learned in unsupervised fashion.

## Results

**A deep generative model for evolutionary sequence data**

One strategy for reasoning about the consequences of mutations to genes is to develop models of the selective constraints that have been relevant throughout evolution. Since the genes that we observe across species today are the results of long-term evolutionary processes that select for functional molecules, a *generative model* of the outputs of evolution must implicitly learn some of these functional constraints. If we approximate the evolutionary process as a "sequence generator" with probability $p(\mathbf{x}|\boldsymbol{\theta})$ that has been fit to reproduce the statistics of evolutionary data, we can use the probabilities that the model assigns to any given sequence as a proxy for the relative plausibility that the molecule satisfies functional constraints. We will consider the log-ratio, $log \frac{p(\mathbf{x}^{(\text{Mutant})}|\boldsymbol{\theta})}{p(\mathbf{x}^{(\text{Wild-Type})}|\boldsymbol{\theta})}$ as a heuristic metric for the relative favorability of a mutated sequence, $\mathbf{x}^{(\text{Mutant})}$, versus a wild-type $\mathbf{x}^{(\text{Wild-Type})}$. This log-ratio heuristic has been previously shown to accurately predict the effects of mutations across multiple kinds of generative models $p(\mathbf{x}|\boldsymbol{\theta})$ [36]. Our innovation is to instead consider another class of probabilistic models for $p(\mathbf{x}|\boldsymbol{\theta})$, nonlinear latent variable models (Figure 1). It is important to emphasize that this new approach, as with the previous pairwise approach, is fully *unsupervised*, as we never train on any observed mutation effect data but rather use the statistical patterns in observed sequences as a signal of selective constraint.

We introduce a nonlinear latent variable model $p(\mathbf{x}|\boldsymbol{\theta})$ to implicitly capture higher order interactions between positions in a sequence in a protein family. For every observed sequence **x,** we posit unobserved latent variables **z** together with a *generative process* p(**z**)p(**x**|**z**) that specifies a joint distribution over hidden variables and observed variables. Inference this model is challenging, as the marginal probability of the observed data, p(**x**), requires integrating over all possible hidden **z** with

$$p(\mathbf{x}|\boldsymbol{\theta}) = \int p(\mathbf{x}|\mathbf{z}, \boldsymbol{\theta})p(\mathbf{z})d\mathbf{z}.$$

While directly computing this probability is intractable in the general case, we can use variational inference[50] to instead form a lower bound on the (log) probability. This bound, known as the Evidence Lower Bound (ELBO), takes the form

$$\log p(x|\boldsymbol{\theta}) \geq \mathbb{E}_q[\log p(\mathbf{x}|\mathbf{z}, \boldsymbol{\theta})] - D_{KL}(q(\mathbf{z}|x, \boldsymbol{\phi})||p(\mathbf{z})),$$

where q(**z**|**x**) is an *approximate posterior* for hidden variables given the observed variables p(**z**|**x**). Modeling both the conditional distribution p(**x**|**z**) of the generative model and the approximate posterior q(**z**|**x**) with neural networks results in a flexible model-inference combination, known as a Variational Autoencoder [47, 48] ( Figure 1b).

Neural network-parameterized latent variable models can in principle model complex correlations in data, but without additional architectural and statistical considerations may be hard to interpret and unlikely to generalize. We encourage generalization in three ways: First, we encourage *sparse interactions* by placing a group sparsity prior over the last layer of the neural network for p(**x**|**z**) that encourages each hidden unit in the network to only influence a few positions at a time. This is motivated by the observation that higher order interactions in proteins, while importantly higher than second order, are nevertheless low-valence compared to the number of residues in the protein. Second, we encourage *correlation between amino acid usage*, by convolving the final layer with a width-1 convolution operation. Thirdly, we estimate all global parameters with variational Bayes by estimating approximate posterior distributions over each model parameter. The result is that rather than learning a single neural network for p(**z**|**x**), we learn an infinite *ensemble* of networks. This joint variational approximation is then optimized by stochastic gradient ascent on the ELBO to give a fully trained model (Methods).

After optimizing the model on a given family, it can be readily applied to predict the effects of arbitrary mutations to arbitrary sequences. Following the previous heuristic of quantifying effects with a log ratio, $log \; \frac{p(\mathbf{x}^{(\text{Mutant})}|\boldsymbol{\theta})}{p(\mathbf{x}^{(\text{Wild-Type})}|\boldsymbol{\theta})}$ , we approximate this quantity by replacing each log probability with a lower bound, the ELBO. For example, given a starting wild type sequence, one can rapidly compute this difference in ELBOs for all possible single point mutations (Figure 2).

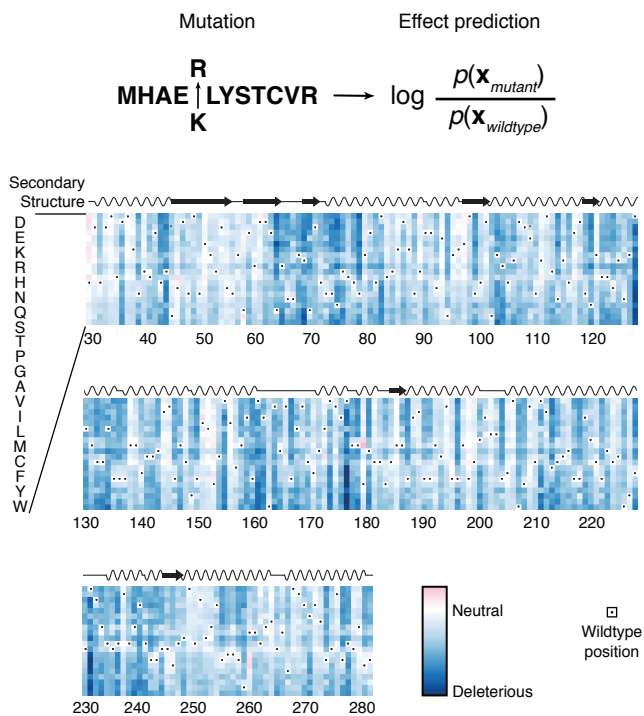

**Figure 2. Mutation effects can be quantified by likelihood ratios.** *After fitting a probabilistic model to a family of homologous sequences, we heuristically quantify the effect of mutation as the log ratio of mutant likelihood to wild type likelihood (as approximated by the ELBO; Methods). Below: mutation effect scores for all possible point mutations to β-lactamase.*

Since all model parameters are computed for any combinations of mutations compared to wild type, sequences can be assessed for fitness that are multiple steps away from the wild type and compared.

**A deep latent variable model captures the effects of mutations**

Deep mutational scanning (DMS) experiments provide a systematic survey of the mutational landscape of proteins and can be used to benchmark computational predictors for the effects of mutations [36]. Here we surveyed 28 *in vivo* and *in vitro* deep mutational scanning experiments comprising of 21 different proteins and a tRNA to assess the ability of the deep latent variable model to predict the effect of mutations purely from natural sequence variation [8, 14, 17, 22, 25, 37, 38, 51-68]. For each multiple sequence alignment of a family, we fit five replicas of the model from 5 different initializations both to assess reproducibility as well as to create an ensemble predictor. We calculate mutation effects as the difference in ELBOs (above and Methods). Our deep latent variable model, DeepSequence, is predictive of the effect of mutations with better performance than a site-independent model without dependencies between positions with the same or better performance in 23 out of 28 datasets) (average 0.110 Spearman ρ increase 0.11, Figure 3a).

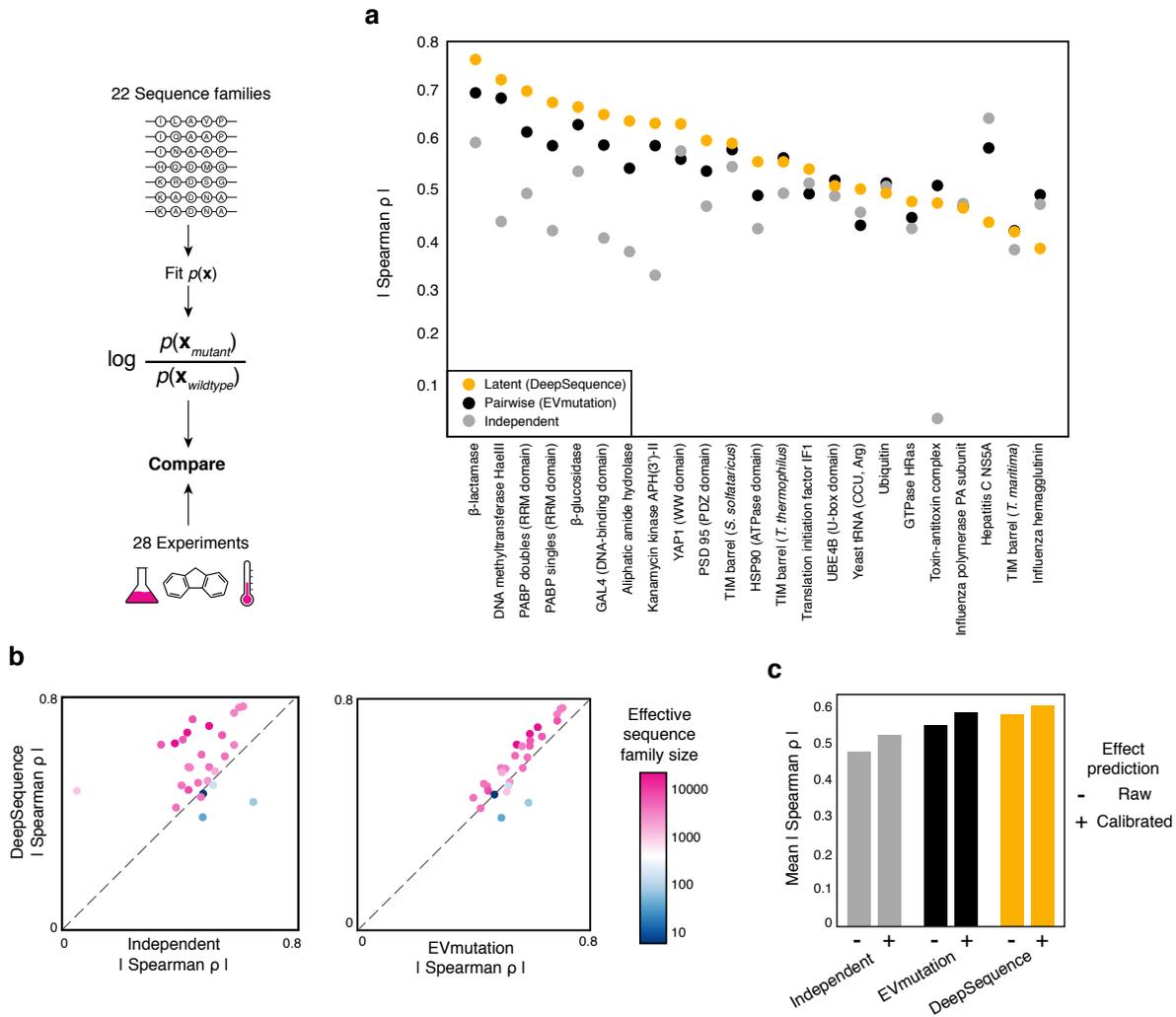

**Figure 3. A deep latent variable model predicts the effects of mutations better than site-independent or pairwise models.** *a. A nonlinear latent variable model (DeepSequence) captures the effects of mutations across deep mutational scanning experiments as measured by rank correlation (Supplemental Figure 1). b. The latent variable model tends to be more predictive of mutational effects than pairwise and site-independent models when fit to deeper, more evolutionarily diverse sequence alignments as measured by the effective family size (Methods). c. Average Spearman ρ before and after bias calibration of representative single-mutant datasets (Methods, Supplementary Figure 3).*

DeepSequence matches or is more predictive than the current state-of-the-art pairwise model in 22 out of 28 datasets (average Spearman ρ increase 0.03) and, as expected, the ensembled prediction of the five model replicas is more predictive than the average performance of individual predictors (28 out of 28 datasets) (Figure 3a, Supplementary Figure 1). A deep alignment is necessary but not sufficient for reasonable agreement between experimental measurements and model predictions. Where the effective family size is greater than 100 sequences, DeepSequence matches or improves predictive performance in 21 out of 24 datasets; in this data regime, the latent variable model increases the average model-data correlation by 0.045. When fit to protein families that have lower effective family sizes (N=4, Neff ($\theta$=0.2) = 92.6, 44.6, 21.8, 3.0), the independent and pairwise model outperform the deep generative model (Figure 3b). We anticipate effective family size can guide model selection for best predictive performance.

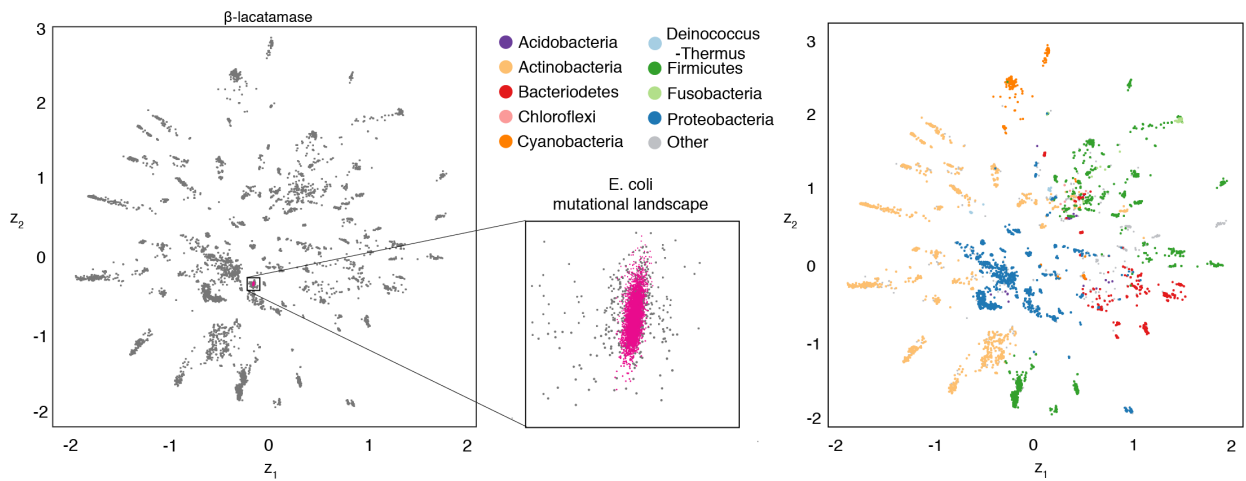

**Figure 4. Latent variables capture organization of sequence space.** *In a two-dimensional latent space for the β-lactamase family, closeness in latent space reflects phylogenetic groupings. When examining the variation within a single deep mutational scanning experiment, it occupies only a very small portion of the sequence space of the entire evolutionary family.*

We compared the residuals of the rankings of the predictions versus the experiment for each amino acid transition and observed a similar prediction bias for all three evolutionary models (independent, EVmutation, DeepSequence, Supplementary Figure 2). When averaged across all possible starting amino acids, positions mutated to prolines and charged amino acids are consistently predicted as too deleterious, while sulphur-containing residues and aromatics are consistently predicted as too fit, (Supplementary Figure 2). Although the unsupervised DeepSequence model presented in this paper is improved by accounting for this bias in datasets with only single mutants, the improvements are small (Figure 3c, Supplementary Figure 3) suggesting that most of the disagreement between DeepSequence and the experimental measurements is more complex.

We found that the combination of using biologically motivated priors and Bayesian approaches for inference on the weights was important to learning models that generalize. To test the importance of these various aspects of the model and inference, we performed an ablation study across a subset of the proteins. We found that using (i) Bayesian variational approximations on the weights, (ii) sparse priors on the last layer, (iii) a final width 1 convolution for amino acid correlations, and (iv) a global temperature parameter all improved the ability of the model to predict the effects of mutations across this subset of the experiments.

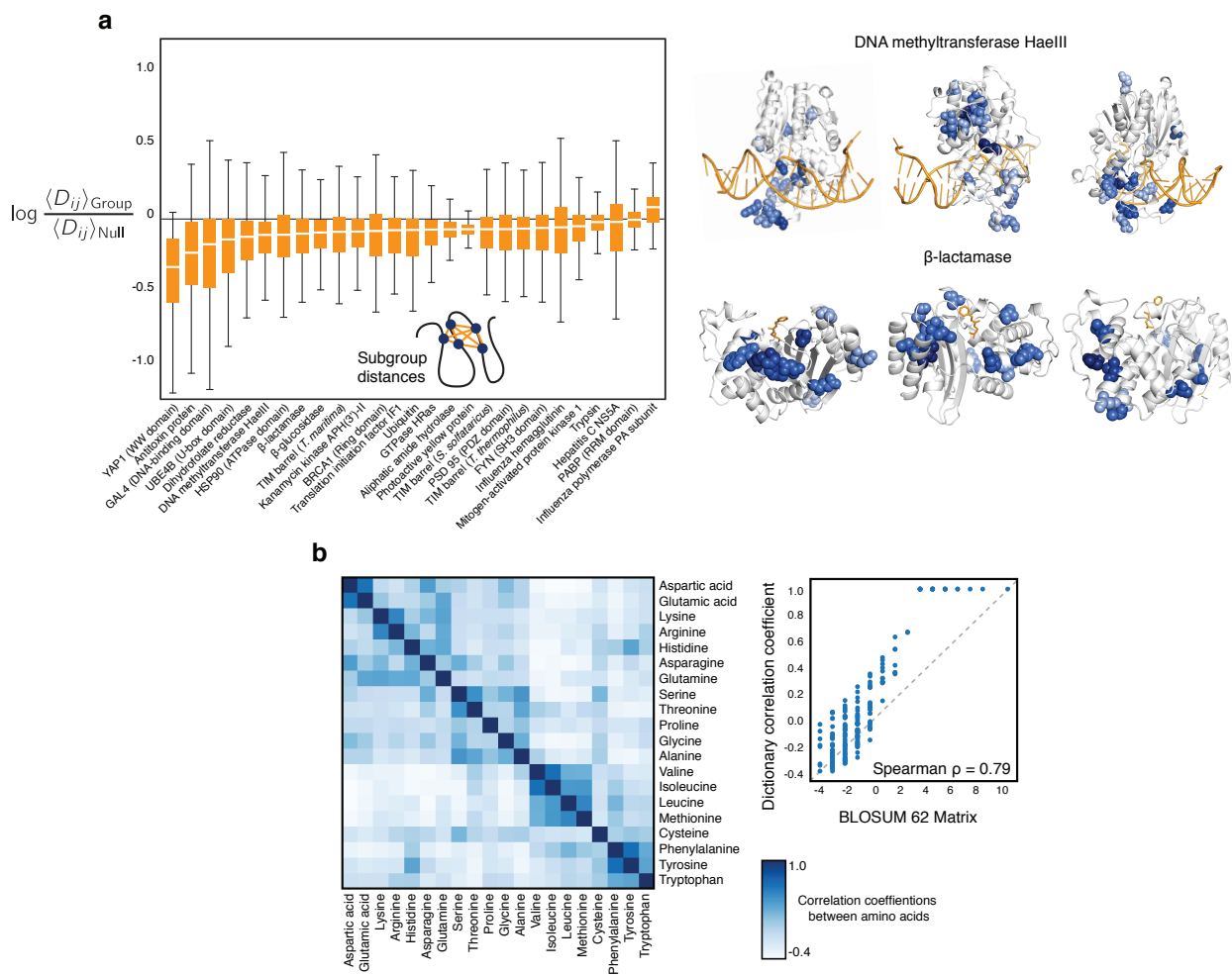

**Figure 5. Model parameters capture structural proximity and amino acid correlations.** *a. (Left) Sparse sub-groups targeted by the latent factors with a group sparsity prior are enriched as closer in structure than expected for random sub-groups. (Right) When visualizing structures with only median values of the structural enrichment, structural proximity is apparent. Representative sparse sub-groups are plotted on DNA methyltransferase HaeIII (pdb: 1dct; log-ratio distance from left to right: -0.11,-0.11,-0.13) and β-lactamase (pdb: 1axb; log-ratio distance from left to right: -0.11,-0.11,-0.10). b. Correlations in the weights of the final width-1 convolution reflect known amino acid correlations as captured by a well-known substitution matrix BLOSUM62 (Spearman ρ = 0.79).*

Moreover, when comparing to other common approaches for regularization such as Dropout [69] or point estimation with group sparsity priors [70], we found that our variational Bayesian approach performed better, (Table 1).Most importantly, only the Bayesian approaches for inference of the global parameters that estimate approximate posteriors were able to consistently outperform the previous pairwise-interaction models.

|  | Bayesian θ | | | | | | MAP θ | | | | | | | Pair | Site |
|---|---|---|---|---|---|---|---|---|---|---|---|---|---|---|---|
| Sparsity [S] | ✓ | ✓ | ✓ |  |  |  |  |  |  |  |  |  |  |  |  |
| Convolution [C] | ✓ | ✓ |  | ✓ |  |  |  |  |  |  |  |  |  |  |  |
| Temperature [T] | ✓ |  | ✓ | ✓ | ✓ |  |  |  |  |  |  |  |  |  |  |
| L2 Regularization |  |  |  |  |  |  | ✓ | ✓ | ✓ | ✓ | ✓ | ✓ |  |  |  |
| Dropout |  |  |  |  |  |  | ✓ | ✓ |  |  | ✓ |  |  |  |  |
| [S+C+T] |  |  |  |  |  |  | ✓ |  | ✓ |  |  |  |  |  |  |
| Final ReLU |  |  |  |  |  |  |  | ✓ |  | ✓ |  |  | ✓ |  |  |
| β-lactamase | 0.73 | 0.73 | 0.73 | 0.73 | 0.73 | **0.74** | 0.53 | 0.61 | 0.04 | 0.40 | 0.56 | 0.37 | 0.34 | 0.42 | 0.70 | 0.60 |
| PSD 95 (PDZ domain) | 0.58 | **0.60** | 0.58 | 0.57 | 0.57 | 0.55 | 0.55 | 0.48 | 0.32 | 0.47 | 0.50 | 0.41 | 0.37 | 0.47 | 0.54 | 0.47 |
| GAL4 (DNA-binding domain) | 0.61 | 0.46 | 0.50 | **0.62** | 0.60 | 0.58 | 0.60 | 0.53 | 0.26 | 0.47 | 0.52 | 0.43 | 0.42 | 0.47 | 0.59 | 0.41 |
| HSP90 (ATPase domain) | **0.54** | **0.54** | **0.54** | 0.51 | 0.52 | 0.52 | 0.48 | 0.45 | 0.03 | 0.34 | 0.44 | 0.26 | 0.22 | 0.33 | 0.49 | 0.43 |
| Kanamycin kinase APH(3')-II | **0.62** | **0.62** | **0.62** | 0.60 | 0.59 | 0.60 | 0.53 | 0.49 | 0.09 | 0.38 | 0.49 | 0.40 | 0.39 | 0.38 | 0.59 | 0.33 |
| DNA methyltransferase HaeIII | **0.70** | **0.70** | 0.69 | **0.70** | 0.68 | 0.68 | 0.64 | 0.64 | 0.12 | 0.54 | 0.64 | 0.50 | 0.49 | 0.54 | 0.69 | 0.44 |
| PABP singles (RRM domain) | **0.67** | **0.67** | 0.66 | 0.65 | 0.63 | 0.65 | 0.64 | 0.62 | 0.44 | 0.59 | 0.63 | 0.58 | 0.58 | 0.59 | 0.59 | 0.42 |
| Ubiquitin | **0.50** | 0.46 | 0.46 | 0.44 | 0.48 | 0.43 | 0.37 | 0.39 | 0.09 | 0.38 | 0.37 | 0.29 | 0.31 | 0.38 | 0.43 | 0.46 |
| YAP1 (WW domain) | **0.64** | **0.64** | **0.64** | 0.63 | 0.63 | **0.64** | 0.63 | 0.58 | 0.28 | 0.50 | 0.61 | 0.49 | 0.44 | 0.50 | 0.57 | 0.58 |

**Table 1. Biologically motivated priors and Bayesian learning improve model performance.**
*Ablation studies of critical components of DeepSequence, showing the average Spearman ρ of predictions from five randomly-initialized models. We include combinations of components of the structured matrix decomposition and use either Bayesian approximation or Maximum a posteriori (MAP) estimation of decoder weights. These can be compared to predictions made from EVmutation (Pair) and the site-independent model (site). Inclusion is indicated with (✓), and top performing model configurations for each dataset are bolded.*

**The latent variables and global variables capture biological structure**

Examining the low-dimensional latent spaces learned by a latent variable model can give insight into relationships between data points (sequences), so we fit an identical replica of the model for Beta-lacatamse that was constrained to have 2-dimensional **z**. We observe that sequence closeness in latent space largely reflects phylogenetic groupings, though with some deviations (Figure 4). Interestingly, when we examine the distribution of single-point mutation sequences in latent, they are tightly clustered. It is important to note that these sequences need not be separated at all; the conditional distribution p(**x**|**z**) can in principle model all of this variation without additional need for variation in latent variables.

For the pairwise model of sequence families, it is well established that strongly coupled positions in the model are also close in protein 3D structure [71-74]. Assessing an analogous pattern in a latent variable model is difficult, however, because explicit correlations between sites in p(**x**) will be implicitly captured by the couplings between observed variables and latent variables. Since these dependencies are mediated by the neural network for p(**x**|**z**) and the observed variables **x** are only directly affected via connections from the last hidden layer, we can focus our attention on those neural network weights. The group sparsity prior over this set of weights (Methods) learns 500 soft sub-groups of positions, which can be seen as subsets of the entire sequence that are jointly influenced by the same hidden activations. We tested if these subgroups tend to be closer in 3D structure than might be expected by chance. For each of these subgroups, we computed the average pairwise distance between positions in the group (after thresholding for inclusion; Methods). We observe that the bulk of these average subgroup distances tends to be less than the null expectation for distance (Figure 5a). When focusing on subgroups with enrichment under the null near the median for that protein, we see that they have many nearby subsets of residues on the crystal structures (Figure 5b). The final width-1 convolution in the network is parameterized by a matrix that captures patterns of amino acid usage. To visualize these patterns, we plot the correlations between amino acid types across the input channels of this matrix and find that it groups amino acids of similar types. Moreover, it is well correlated with the widely used BLOSUM62 substituion matrix (Figure 5c).

## Discussion

We have shown that a deep latent variable model can model variation in biological sequence families and be applied to predict the effects of mutations across diverse classes of proteins and RNAs. We find that the predictions of the deep latent variable model are more accurate than a previously published pairwise-interaction approach to model epistasis [36, 75], which in turn was more accurate than commonly used supervised methods [76, 77]. In addition, both the latent variables and global variables of the model learn interpretable structure both for macrovariation and phylogeny as well as structural proximity of residues.

However, while deep latent variable models introduce additional flexibility to model higher-order constraints in sequence families, this comes at the price of reduced interpretability and increased potential for overfitting. We find that a Bayesian approach to inference, where averages are computed over an ensemble of models and global parameters are controlled by group sparsity priors, was a crucial step towards attaining generality. This suggests that future work could benefit from additional biologically-motivated, hierarchical priors as well as more accurate methods for variational inference [78, 79]. Additionally, incorporating more rigidly structured probabilistic graphical models to model dependencies between latent variables could improve generality and interpretability [80]. Even our preliminary results with group sparsity priors suggest that fear of a tradeoff between interpretability and flexibility for using deep models on biological data may be largely remedied by hierarchical Bayesian approaches for modeling.

A second challenge for all approaches that predict the effects of mutations from evolutionary sequence variation concerns the data themselves. DeepSequence, as with the majority of previous mutation prediction methods, rely critically on the multiple sequences alignments used for training data [36, 72, 81-83]. At present, the criteria for the numbers of non-redundant sequences and the level of diversity in multiple sequence alignments is ad hoc and this should be improved to give better uncertainty criteria and accuracy expectation in predicting the effects of mutations Secondly, the premise that evolutionary data can be applied to predict outcomes of an experiment is highly contingent on the relevance of the experimental assay to long-term selective forces in the family. A mutation may be damaging with regard to some measurable protein feature e.g. enzyme efficiency, but harmless for stability or even organism fitness, as we and others have previously discussed [36, 38, 63]. We therefore suggest that DeepSequence could be incorporated into umbrella or supervised methods to enhance prediction for specific purposes such as disease risk, binding specificity or enzyme efficiency.

Despite challenges for deep models of sequence variation and data used to train them, they are likely to be of increasing importance to the high-throughput design and annotation of biological sequences. Evolution has generated millions of protein experiments, and deep generative models can begin to identify the statistical patterns of constraint that characterize essential functions of molecules. We make both the model and datasets available at github.com/debbiemarkslab/DeepSequences

**Acknowledgements**
We thank Chris Sander, Frank Poelwijk, David Duvenaud, Sam Sinai, Eric Kelsic and members of the Marks lab for helpful comments and discussions. While in progress Sinai et al also reported on use of variational autoencoders for protein sequences [84]. A.J.R. is supported by DOE CSGF fellowship DE-FG02-97ER25308. D.S.M. and J.B.I. were funded by NIGMS (R01GM106303)

## Methods

**Alignments.**

We used the multiple sequence alignments that were published with EVmutation for the 19 families that overlapped [36] and repeated the same alignment-generation protocol for the 4 additional proteins that were added in this study. Briefly, for each protein (target sequence), multiple sequence alignments of the corresponding protein family were obtained by five search iterations of the profile HMM homology search tool jackhmmer[85] against the UniRef100 database of non-redundant protein sequences[86] (release 11/2015). We used a bit score of 0.5 bits/residue as a threshold for inclusion unless the alignment yielded < 80% coverage of the length of the target domain, or if there were not enough sequences (redundancy-reduced number of sequences $\geq 10L$). For $<10L$ sequences, we decreased the required average bit score until satisfied and when the coverage was < 80% we increased the bit score until satisfied. Proteins with < 2L sequences at < 70% coverage were excluded from the analysis. See previous work for ParE-ParD toxin-antitoxin and tRNA alignment protocols.

**Sequence weights.** The distributions of protein and RNA sequences in genomic databases are biased by both (i) human sampling, where the sequences of certain highly-studied organisms may be overrepresented, and (b) evolutionary sampling, where some types of species may have undergone large radiations that may not have anything to do with the particular molecule we are studying. We aim to reduce these biases in a mechanistically-agnostic way by reweighting the empirical data distribution to make it smoother. We use the previously established procedure of computing each [87] sequence weight $\pi_s$ as the reciprocal of the number of sequences within a given Hamming distance cutoff. If $D_H(X^s, X^t)$ is the normalized hamming distance between the query sequence $X^s$ and another sequence in the alignment $X^T$ and $\theta$ is a pre-defined neighborhood size, the sequence weight is:

$$\pi_s = \left( \sum_t^T I[D_H(X^s, X^t) < \theta] \right)^{-1}$$

The effective sample size of a multiple sequence alignment can then be computed as the sum of these weights as

$$N_{eff} = \sum_t^T \pi_t$$

To fit a model to reweighted data, there are two common approaches. First, as was done previously[87], one can reweight every log-likelihood in the objective by its sequence weight $\pi_s$. While this works well for batch optimization, we found it to lead to high-variance gradient estimates with mini-batch optimization that make stochastic gradient descent unstable. We instead used the approach of sampling data points with probability $p_s$ proportional to their weight in each minibatch as

$$p_s = \frac{\pi_s}{N_{eff}}$$

Following prior work[36], we set $\theta = 0.2$ for all multiple sequence alignments sequences (80%sequence identity) except those for viral proteins where we set $\theta = 0.01$ (99% sequence identity) due to limited sequence diversity and the expectation that small differences in viral sequences have higher probability of containing constraint information that the same diversity might from a sample of mammals, for instance.

**Background: latent factor models.** Probabilistic latent variable models reveal structure in data by positing an unobserved generative process that created the data and then doing inference to learn the parameters of the generative process. We will focus on models with a generative process in which an unobserved set of factors **z** are drawn from an in-dependent distribution and each data point arises according to a conditional distribution $p(\mathbf{x}|\mathbf{z}, \boldsymbol{\theta})$ that is parameterized by $\boldsymbol{\theta}$. This process can be written as

$$z \sim \mathcal{N}(0, I_D)$$
$$x \sim p(x|z, \theta)$$

Principal Component Analysis (PCA) has been a foundational model for the analysis of genetic variation since its introduction by Cavalli-Sforza. PCA can be realized in this probabilistic framework as the zero-noise limit of Probabilistic PCA[44, 88]]. With linear conditional dependencies $p(\mathbf{x}|\mathbf{z}, \boldsymbol{\theta})$, PCA can only model additive interactions between the latent factors $\mathbf{z}$. This limitation could in principle be remedied by using a conditional model $p(\mathbf{x}|\mathbf{z}, \boldsymbol{\theta})$ with nonlinear dependencies on $z$.

**Nonlinear categorial factor model.** We will consider a conditional model $p(\mathbf{x}|\mathbf{z}, \boldsymbol{\theta})$ that differs from PCA in two ways: First, the conditional distribution of the data $p(\mathbf{x}|\mathbf{z}, \boldsymbol{\theta})$ will be categorical rather than Gaussian to model discrete characters. Second, the conditional distribution $p(\mathbf{x}|\mathbf{z}, \boldsymbol{\theta})$ will be parameterized by a neural network rather than a linear map. In this sense, our latent variable model may be thought of as a discrete, nonlinear analog of PCA.

For this work, we considered a simple two hidden-layer neural network parameterizations of $p(\mathbf{x}|\mathbf{z}, \boldsymbol{\theta})$. The generative process $p(\mathbf{x}|\mathbf{z}, \boldsymbol{\theta})$ specifying the conditional probability of letter $a$ a at position $i$ is

$$z \sim \mathcal{N}(0, I_D)$$
$$h^{(1)} = f_1(W^{(1)}z + b^{(1)})$$
$$h^{(2)} = f_2(W^{(2)}h^{(1)} + b^{(2)})$$
$$h^{(3,i)} = W^{(3,i)}h^{(2)} + b^{(3,i)}$$
$$p(x_i = a|z) = \frac{exp(h_a^{(3,i)})}{\sum_b exp(h_b^{(3,i)})}$$

where $f_1 = \max(0, u)$ and $f_2 = \frac{1}{1+e^{-u}}$.

**Structured sparsity**. Motivated by the observation that sequences have been well described by models with low-order interactions (such as pairwise undirected models), we structure the final layer of the decoder to be sparse such that each hidden unit may only affect a few positions in the sequence. We parameterize each final weight matrix as

$$W^{(3,i)} = \log(1 + e^\lambda) \frac{1}{1 + e^{-s_{i\,mod\,(\frac{H}{K})}}} \widetilde{W}^{(3,i)} D$$

where $\frac{1}{1+e^{-s_{i\,mod\,(\frac{H}{k})}}}$ is a sigmoid function representing a continuous relaxation of a spike and slab prior over the group of dense factors using a logit normal prior. A single set of scale parameters can control the sparsity of $k$ dense factors out of the total number factors $H$ by tiling. $\log(1 + e^\lambda)$ is a softmax function representing the inverse-temperature of the sequence family and $D$ is the dictionary.

The priors over the decoder weights are:

$$\widetilde{W} \sim \mathcal{N}(0, I)$$
$$D \sim \mathcal{N}(0, I)$$
$$s \sim \mathcal{N}(\mu_s, \sigma_s^2)$$
$$\lambda \sim \mathcal{N}(0, I)$$

The factors $S = \frac{1}{1+e^{-s_{i\,mod\,(\frac{H}{k})}}}$ are a-priori logit-normally distributed, which can be though of as a smooth relaxation of a Bernoulli that can be made sharper by increasing the variance $\sigma_s^2$, We set $\mu_s$ such that the prior probability of approximate inclusion, $P\left(\frac{1}{1+e^{-s_{i\,mod\,(\frac{H}{k})}}} > 0.5\right)$, was 0.01. Given a fixed logit-variance of $\sigma_s^2 = 16$ and an inclusion probability $p_{include} = 0.01$, we set the prior mean for the logit as $\mu_s = -9.3053$ using

$$\mu_s = \sqrt{2\sigma_s^2}\,erf^{-1}(2p_{include} - 1)$$

**Variational approximation to $p(\mathbf{z}|\mathbf{x}, \boldsymbol{\theta})$.** Nonlinear latent factor models are difficult to infer. Since the latent variables $z$ are not observed, computing the marginal likelihood of the data requires integrating them out:

$$\log p(\mathbf{x}|\boldsymbol{\theta}) = \log \int p(\mathbf{x}|\mathbf{z}, \boldsymbol{\theta})\, p(\mathbf{z})d\mathbf{z}$$

We must do this integral because we do not know *a priori* which $\mathbf{z}$ is responsible for each data point $\mathbf{x}$, and so we average over all possible explanations weighted by their relative probability. In principle, the conditional probability $p(\mathbf{z}|\mathbf{x}, \boldsymbol{\theta})$ is given by Bayes' Theorem as the posterior,

$$p(\mathbf{z}|\mathbf{x}, \boldsymbol{\theta}) = \frac{p(\mathbf{x}, \mathbf{z}|\boldsymbol{\theta})}{p(\mathbf{x})} = \frac{p(\mathbf{x}|\mathbf{z}, \boldsymbol{\theta})p(\mathbf{z})}{\int p(\mathbf{x}|\mathbf{z}, \boldsymbol{\theta})p(\mathbf{z})\,d\mathbf{z}},$$

which is a challenging calculation that requires integrating over $\mathbf{z}$.

Kingma and Welling [47], and Rezende et al., [48] showed how to tractably overcome the challenge of the intractable posterior by introducing a variational approximation $q(\mathbf{z}|\mathbf{x}, \boldsymbol{\phi})$. By Jensen's inequality, this forms a lower bound on the marginal likelihood of a data point $\mathbf{x}$ as

$$\log p(\mathbf{x}|\boldsymbol{\theta}) = \log \int p(\mathbf{x}|\mathbf{z}, \boldsymbol{\theta})p(\mathbf{z})d\mathbf{z}$$
$$= \log \int \frac{p(\mathbf{x}|\mathbf{z}, \boldsymbol{\theta})p(\mathbf{z})}{q(\mathbf{z}|\mathbf{x}, \boldsymbol{\phi})} q(\mathbf{z}|\mathbf{x}, \boldsymbol{\phi})d\mathbf{z}$$
$$\geq \int \log \frac{p(\mathbf{x}|\mathbf{z}, \boldsymbol{\theta})p(\mathbf{z})}{q(\mathbf{z}|\mathbf{x}, \boldsymbol{\phi})} q(\mathbf{z}|\mathbf{x}, \boldsymbol{\phi})d\mathbf{z}.$$

We can write this lower bound as:

$$\log p(\mathbf{x}|\boldsymbol{\theta}) \geq \mathbb{E}_q[\log p(\mathbf{x}|\mathbf{z}, \boldsymbol{\theta})] - D_{KL}\big(q(\mathbf{z}|\mathbf{x}, \boldsymbol{\phi})||p(\mathbf{z})\big)$$

We choose the following functional form for the variational approximation for $\mathbf{z}$:

$$g^{(1)} = f_1\big(W^{(1)}\mathbf{x} + b^{(1)}\big)$$
$$g^{(2)} = f_1\big(W^{(2)}g^{(1)} + b^{(2)}\big)$$
$$\mu = W_\mu^{(3)}g^{(2)} + b_\mu^{(3)}$$
$$\sigma^2 = W_\sigma^{(3)}g^{(2)} + b_\sigma^{(3)}$$
$$q(\mathbf{z}|\mathbf{x}, \boldsymbol{\phi}) = \mathcal{N}(\mu, diag(\sigma^2))$$

The latent variable $\mathbf{z}$ can be reparameterized using an auxillary random variable $\epsilon \sim \mathcal{N}(0, I)$:

$$z = \mu + \sigma * \epsilon$$

**Variational approximation to $p(\boldsymbol{\theta}|X)$.** We apply a Bayesian approach to learning global parameters by extending the variational approximations to include both the latent variables z as well as the global parameters $\boldsymbol{\theta}$. Because the posterior for the global parameters is conditioned on the entire dataset, we must consider the marginal likelihood of the full dataset $X = \{\mathbf{x}^{(1)}, \dots, \mathbf{x}^{(N)}\}$ which integrates out all the corresponding latent factors $Z = \{\mathbf{z}^{(1)}, \dots, \mathbf{z}^{(N)}\}$

$$\log p(X) = \log \iint p(X|Z, \boldsymbol{\theta})p(\boldsymbol{\theta})p(Z)dZd\boldsymbol{\theta}$$

$$= \log \iint \frac{p(X|Z, \boldsymbol{\theta})p(Z)p(\boldsymbol{\theta})}{q(Z, \boldsymbol{\theta}|X, \boldsymbol{\phi})} q(Z, \boldsymbol{\theta}|X, \boldsymbol{\phi})dZd\boldsymbol{\theta}$$

$$\geq \mathcal{L}(\boldsymbol{\theta}, \boldsymbol{\phi}) \triangleq \iint \log \frac{p(X|Z, \boldsymbol{\theta})p(Z)p(\boldsymbol{\theta})}{q(Z, \boldsymbol{\theta}|X, \boldsymbol{\phi})} q(Z, \boldsymbol{\theta}|X, \boldsymbol{\phi})dZd\boldsymbol{\theta}$$

The variational approximation factorizes as

$$q(\mathbf{Z}, \boldsymbol{\theta}|\mathbf{X}, \boldsymbol{\phi}) = q(\mathbf{Z}|\mathbf{X}, \boldsymbol{\phi})q(\boldsymbol{\theta}|\boldsymbol{\phi})$$

The approximate posterior for **Z** factorizes over the data

$$q(\mathbf{Z}|\mathbf{X}, \boldsymbol{\phi}) = \prod_i q(\mathbf{z}^{(i)}|\mathbf{x}^{(i)}, \boldsymbol{\phi})$$

The approximate posterior for factorizes over the model parameters:

$$q(\boldsymbol{\theta}|\boldsymbol{\phi}) = \prod_i (\boldsymbol{\theta}^{(i)}|\boldsymbol{\phi})$$

To incorporate both of these factors into the likelihood, the ELBO is then:

$$\log p(\mathbf{X}|\boldsymbol{\theta}) \geq N \mathbb{E}_{x \in X}\left[\mathbb{E}_{q(\boldsymbol{\theta})q(\mathbf{z}|\mathbf{x})}[\log p(\mathbf{x}|\mathbf{z}, \boldsymbol{\theta})] - D_{KL}(q(\mathbf{z}|\mathbf{x}, \boldsymbol{\phi})||p(\mathbf{z}))\right]$$
$$- \sum_{\boldsymbol{\theta}^{(i)}} D_{KL}\left(q(\boldsymbol{\theta}^{(i)})||p(\boldsymbol{\theta}^{(i)})\right)$$

We model all variational distributions over the parameters with fully-factorized mean-field Gaussian distributions. In accordance with our data reweighting scheme, we set $N = N_{eff}$, the effective number of sequences that is the sum of the sequence weights.

**Model hyperparameters.**

We used a fixed architecture across all sequence families. The encoder has architecture 1500-1500-30 with fully connected layers and ReLU nonlinearities. The decoder has two hidden layers: the first with size 100 and a ReLU nonlinearity, and the second with size 2000 with a sigmoid nonlinearity. The dictionary D is a 40 by q matrix where the alphabet size q was 20 for proteins and 4 for nucleic acids. A single set of sparsity scale parameters controlled 4 sets of dense weights. Dropout [69] was set to 0.5 when used in ablation studies. Models were optimized with Adam [89] with default parameters using a batch size of 100 until convergence, completing 300000 updates.

Each model was fit five times to the same multiple sequence alignment using a different random seed. For mutation effect prediction, 2000 ELBO samples were taken of a mutated and wildtype sequence and averaged to estimate the log probability of that mutation.

**Group sparsity analysis.** The sparse scale parameters were introduced into the structured weight matrix decomposition to enable the model to capture low-valence interactions between residues. We aimed to test if positional inclusions for a given hidden unit were closer in three dimensions than expected by chance.

To gather ground-truth distance data from protein structures, we retrieved related structures by searching the 3D structure database [90] with our query sequences ui9sng Jackhmmer. We computed multiple distance matrice $D$ were generated by taking both the median of the minimum atom distances of both intra and multimer contacts. The final distance matrix was generated by taking the minimum of both of these matrices.

The approximate sparsity parameters in our network  We use the median of the scale parameters to approximate the value of the scale parameter. For a given vector of scale parameters activated by a neuron $k$, the median activity of a given vector of scale parameters $s_k$:

$$s_k = \frac{1}{1 + e^{-\mu_k}}$$

Since the upstream activation is a sigmoid nonlinearity ($\in [0,1]$), we denoted these dead scale parameter vectors $s_k$ as those which do not have any scale parameter above 0.001, and were removed from downstream analysis.

We then determined a co-occurrence distance distribution of these scale parameters by first taking the upper triangle of the outer product of the scale parameters and normalizing it such that it sums to 1:

$$S_{ijk} = \frac{S_{ik} S_{jk}}{\sum_i^L \sum_j^i S_{ik} S_{jk}} \quad i < L, j < i$$

A normalized distance per vector of scale parameters $D_k^{pattern}$ can then be reported in Angstroms:

$$D_k^{pattern} = \sum_i^L \sum_j^i D_{ij} S_{ijk}$$

This value was compared to $D_{null}^{pattern}$, in which the null distribution of scale parameters $S_{null}$ are isotropically distributed, generating a $D_{null}^{pattern}$ which is the average pairwise distance between residues. Moreover, bootstrapped samplings of $S_k$ converge to the same null value. The distribution of all distances $D^{pattern}$ can be compared to the using a one-sided Student's t-test with a known mean.

**Residual analysis**. Spearman ρ is calculated by transforming paired data to ranked quantiles and then computing the Pearson correlation between the ranks. To determine where the model over or under predicted the ΔE for each mutation, we transformed the experimental measurements and mutation effect predictions to normalized ranks on the interval [0,1]. Thus, we define the residual effects as the residuals of a least-squares linear fit between the normalized ranks.

least-squared line was fit between the normalized ranks of the predictions $d_{\Delta E}$ to the normalized ranks of the experiments $d_{\text{experiment}}$, creating a slope and bias parameter, $m$ and $b$, respectively. Residuals were generated from the fit line:

$$\varepsilon_{\Delta E} = (md_{\Delta E} + b) - d_{\text{experiment}}$$

Positive $\varepsilon_{\Delta E}$ values represent underprediction of deleteriousness of experimental effect prediction, while negative $\varepsilon_{\Delta E}$ values represent overprediction of deleteriousness. Deep mutational scans with only single mutations were analyzed, using the most recent experimental data for each protein. Residuals were grouped by the identity of the amino acid either before mutation (wildtype) or after mutation (mutant).

**Bias correction.** To correct for biases between mutation effect predictions and experimental measurements, we created a feature matrix for each mutation that included ΔE, amino acid identity before and after mutation, alignment column statistics (conservation and amino acid frequency), and residue hydrophobicity[91]. Leave-one-out cross validation (LOOCV) was used to correct the bias for each dataset. Using the most recent DMS experiment as the representative of that protein family (15 DMS datasets), the mutants of 14 datasets were used to fit a regression model to predict the residuals of each known mutation, $\varepsilon_{\Delta E}$, given the feature matrix. After this model was fit, it was used to predict $\varepsilon_{\Delta E}$ for the mutants in the test dataset. This predicted residual bias $\hat{\varepsilon}_{\Delta E}$ was subtracted off the normalized predicted rank $\hat{d}_{\Delta E} = d_{\Delta E} - \hat{\varepsilon}_{\Delta E}$. These corrected predictions were then reranked and compared to the experimental results to calculate a corrected Spearman ρ. To predict the effects of mutations solely from DMS data, the same LOOCV procedure was used excluding all evolutionary information in the feature matrix for each mutation. In this case, the feature matrix was used to directly compute predict a rank $\hat{d}_{\text{DMS}}$. These values were subsequently reranked and compared to the ranked experimental results to calculate a corrected Spearman ρ.

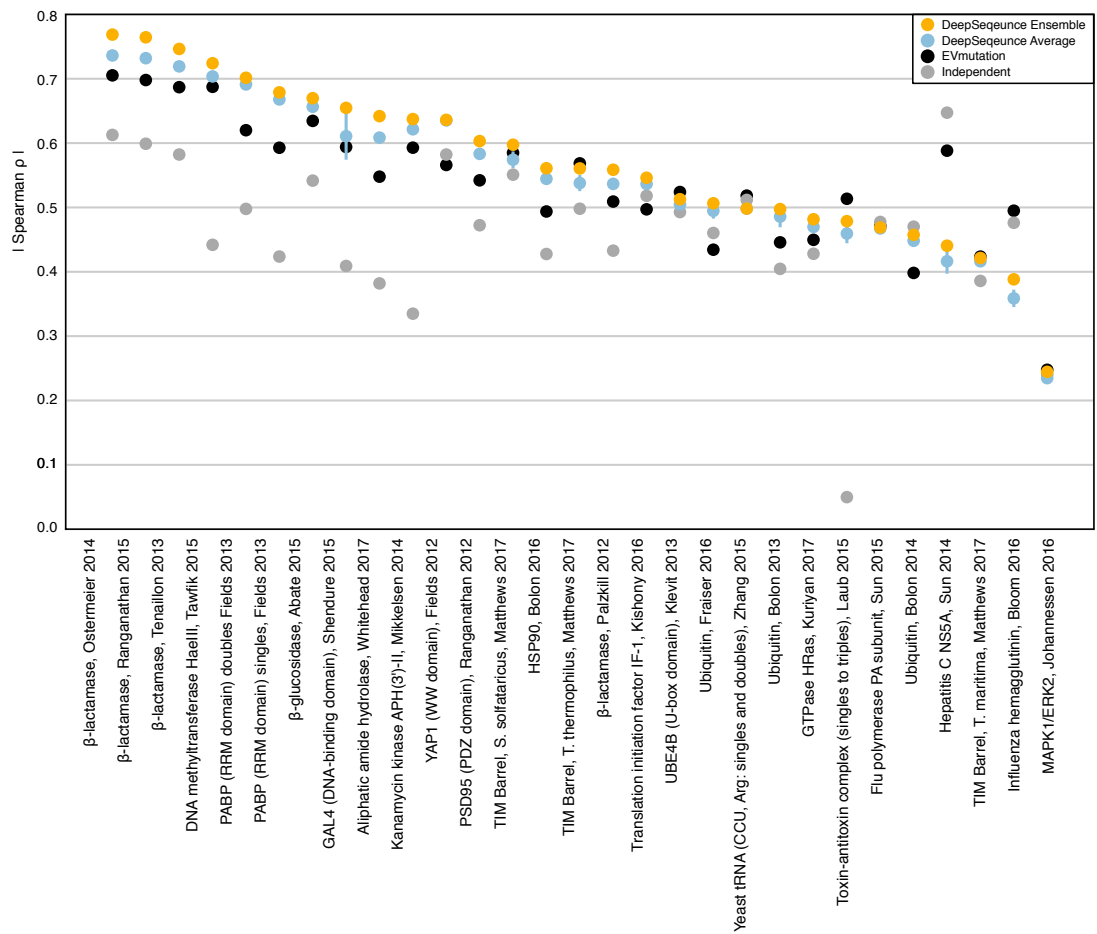

**Supplementary Figure 1. Mutation effect predictions of all deep mutational scanning datasets.** Spearman rank correlation between coefficients of all proteins and all generative models. Here we show both the average rank correlation of individual latent variable models (DeepSequence Average) as well as an ensembled prediction using these 5 models (DeepSequence Ensemble).

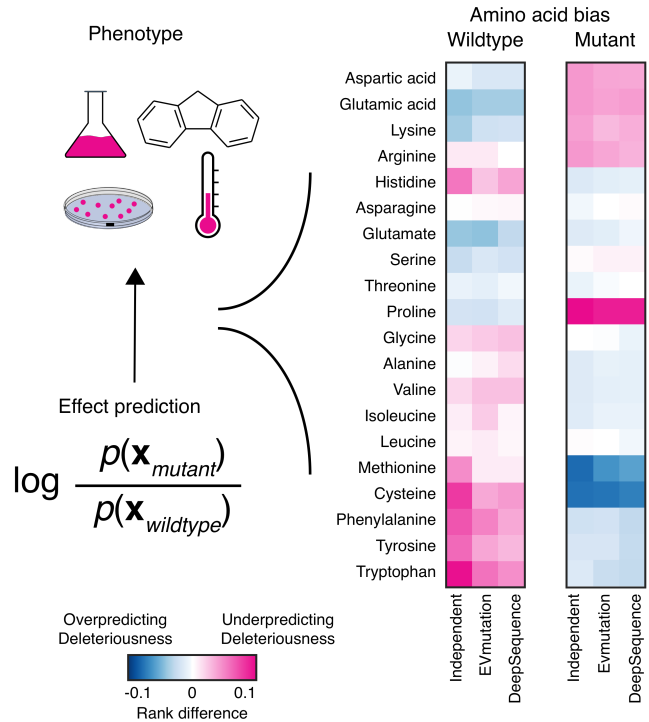

**Supplementary Figure 2. Predictions from all generative models for sequence families exhibit biases when compared to experiments.** By transforming all model predictions and mutations to normalized ranks, we can compare effect predictions to experimental data across all biological datasets and models. The site-independent, pairwise, and latent variable models systematically over and under predict the effects of mutations according to amino acid identity. These biases vary in magnitude and direction depending on the amino acid identity before mutation (wildtype) or the residue identity it is mutated to (mutant).

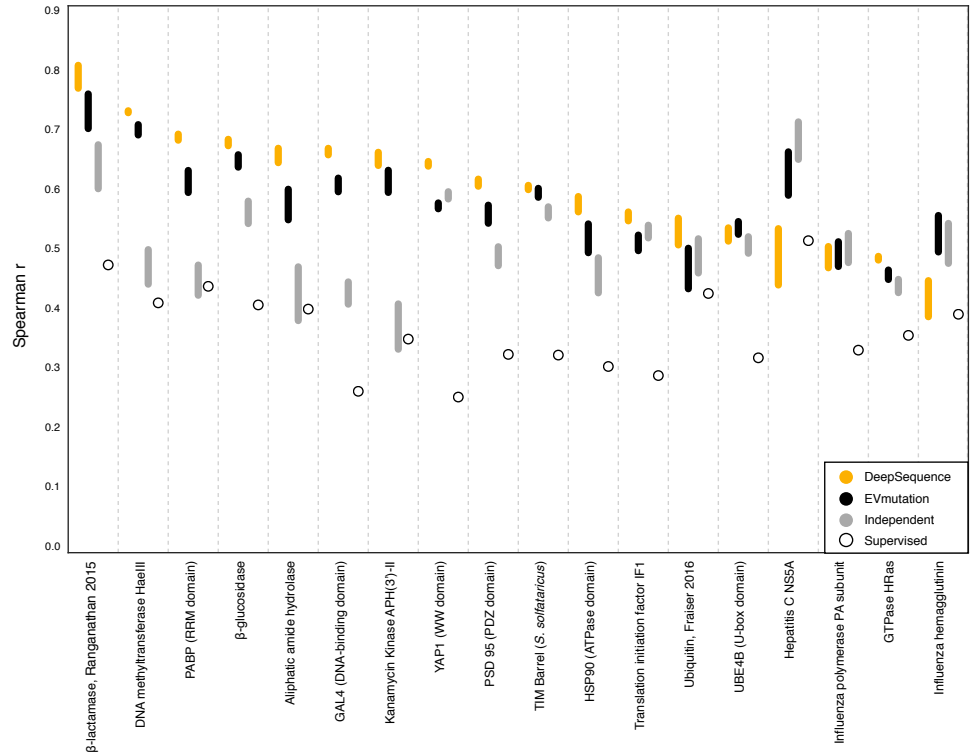

**Supplemental Figure 3. Supervised calibration of mutation effect predictions improves predictive performance.** Amino acid bias was corrected with linear regression for all generative models, leaving one protein out for test and training a model on the rest (Methods). The bottom of the bar is Spearman ρ before correction, while the top is Spearman ρ after correction. Predictions without any evolutionary information (Supervised) performed considerably worse than other predictors.